\documentclass[final,5p,times,twocolumn,authoryear]{elsarticle}

\usepackage{amssymb}
\usepackage{lipsum}
\usepackage{siunitx}
\usepackage{tabularx}
\usepackage{array}
\usepackage{lineno}

\journal{Astronomy $\&$ Computing}

\begin{document}

\begin{frontmatter}



\title{Position reconstruction using deep learning for the HERD PSD beam test}
\cortext[cor1]{Corresponding author.}
\author[label1,label2]{Longkun Yu}
\author[label1,label3]{Chenxing Zhang}
\author[label3]{Dongya Guo\corref{cor1}}
\ead{guody@ihep.ac.cn}
\author[label3]{Yaqing Liu}
\author[label3]{Wenxi Peng\corref{cor1}}
\ead{pengwx@ihep.ac.cn}
\author[label3]{Zhigang Wang}
\author[label3]{Bing Lu}
\author[label3]{Rui Qiao}
\author[label3]{Ke Gong}
\author[label1]{Jing Wang}
\author[label1]{Shuai Yang}
\author[label1]{Yongye Li}


\affiliation[label1]{organization={School of Information Engineering, Nanchang University},
            city={Nanchang},
            postcode={330031},
            country={China}}
\affiliation[label2]{organization={School of Advanced Manufacturing, Nanchang University},
            city={Nanchang},
            postcode={330031},
            country={China}}
            
\affiliation[label3]{organization={Key Laboratory of Particle Astrophysics, The Institute of High Energy Physics of the Chinese Academy of Sciences},
            city={Beijing},
            postcode={100049}, 
            country={China}}

\begin{abstract}
The High Energy cosmic-Radiation Detection (HERD) facility is a dedicated high energy astronomy and particle physics experiment planned to be installed on the Chinese space station, aiming to detect high-energy cosmic rays (\si{\giga\electronvolt} $\sim$ \si{\peta\electronvolt}) and high-energy gamma rays ($> \SI{500}{\mega\electronvolt}$). The Plastic Scintillator Detector (PSD) is one of the sub-detectors of HERD, with 
its main function of providing real-time anti-conincidence signals for gamma-ray detection and the secondary function of measuring the charge of cosmic-rays. In 2023, a prototype of PSD was developed and tested at CERN PS\&SPS. In this paper, we investigate the position response of the PSD using two reconstruction algorithms: the classic dual-readout ratio and the deep learning method (KAN \& MLP neural network). With the latter, we achieved a position resolution of 2 mm ($1\sigma$), which is significantly better than the classic method.
\end{abstract}


\end{frontmatter}


\section{Introduction}
\label{introduction}
\begin{figure}[!htb]
	\centering 
	\includegraphics[width=0.4\textwidth, angle=0]{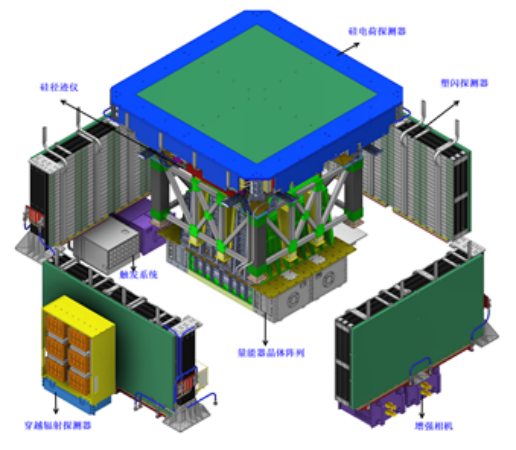}	
	\caption{Layout diagram of HERD} 
	\label{fig1}%
\end{figure}

The High Energy cosmic-Radiation Detection facility (HERD) is a large-scale space particle detection experiment project led by the Institute of High Energy Physics, Chinese Academy of Sciences\citep{zhang2014high}. Its core scientific objectives include precise measurement of cosmic-ray electron spectrum and search for dark matter signals, as well as research on the origin, acceleration, and propagation mechanisms of cosmic-rays, while also conducting all-sky surveys and monitoring of high-energy gamma rays\citep{2021The}\citep{2023Gamma}. HERD project plans to operate for over 10 years on the Chinese space station, with its core scientific capabilities expected to maintain a significant international lead for a long time\citep{2023Latest}. HERD adopts innovative design with three-dimensional position sensitive detectors, as shown in Figure \ref{fig1}, mainly consisting of the 3D-CALOrimeter (CALO), Silicon Charge Detector (SCD), Plastic Scintillator Detector (PSD), Silicon Tracker (STK), and Transition Radiation Detector  (TRD)\citep{kyratzis2020herd}\citep{Kyratzis_2022}\citep{2019The}.

The Plastic Scintillator Detector (PSD) acts primarily as an anti-coincidence detector which helps reject the in-orbit charge particles; meanwhile, it can also perform charge measurement.\citep{gargano2019plastic}\citep{2019Ion}\citep{alemanno2022preliminary}.  The PSD is located on the inner side of SCD, covering the the top and four lateral surfaces of CALO. Due to the size constraints of HERD,  only one layer of plastic scintillators can be accommodated. In order to meet the coverage requirement, segmented plastic scintillator bars with trapezoidal and parallelogram cross-sections are combined to achieve a tight assembly. Each PSD bar has a thickness of \SI{1}{\centi\meter}, a length of approximately \SI{45}{\centi\meter}, and a width of  ranging from \SI{3}{\centi\meter} to $\sim$ \SI{5}{\centi\meter}, depending on the cross-sectional shape. To minimize the gap between the bars, the SiPMs arrays can only be coupled to the upper surface of bars for scintillation light readout\citep{2015Development}\citep{2020Particle}.
Thanks to the bar shape of the PSD, the readout signal of the PSD will be sensitive to the hit positions of particles, which can be reconstructed to help the SCD and STK eliminate ghost hits and improve tracking reconstruction accuracy.\citep{ALEMANNO2023168237}.
In 2023, we tested our first prototype of the PSD at CERN. This paper will focus on studying the position reconstruction algorithm and presenting the analysis results based on the beam test data.

\section{Prototype of PSD}
\begin{figure}
	\centering 
	\includegraphics[width=0.4\textwidth, angle=0]{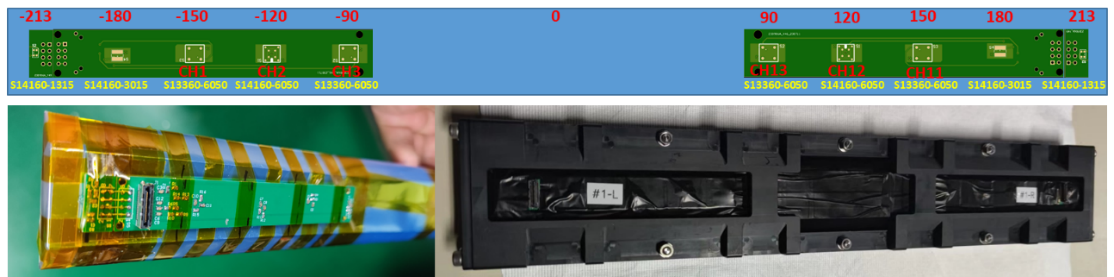}	
	\caption{PSD sample prototype} 
	\label{fig2}%
\end{figure}

The PSD prototype utilizes two trapezoidal cross-section plastic scintillator bars, which are encapsulated with ESR film. As shown in Figure \ref{fig2}, the upper surface of each plastic scintillator is coupled with two arrays of SiPMs. 
Each array has five Hamamatsu MPPCs(SiPMs), which includes two kinds of pixel sizes (50 $\mu m$ and 15 $\mu m$) . Three large pixel-sized SiPMs (one S14160-6050 and two S13360-6050) are mainly used for generating veto signals of all kinds of charge particles, while two small pixel-sized SiPMs (one S14160-3015 and one S14160-1315) are used to measure the charge of cosmic-rays.

In addition, the PSD prototype is powered by a DC-stabilized power supply and a power conversion board and utilizes readout electronics based on CITIROC chips to achieve PSD triggering and data acquisition functions. Scanning tests at different positions are carried out through the displacement platform.

\begin{figure}
	\centering 
	\includegraphics[width=0.5\textwidth, angle=0]{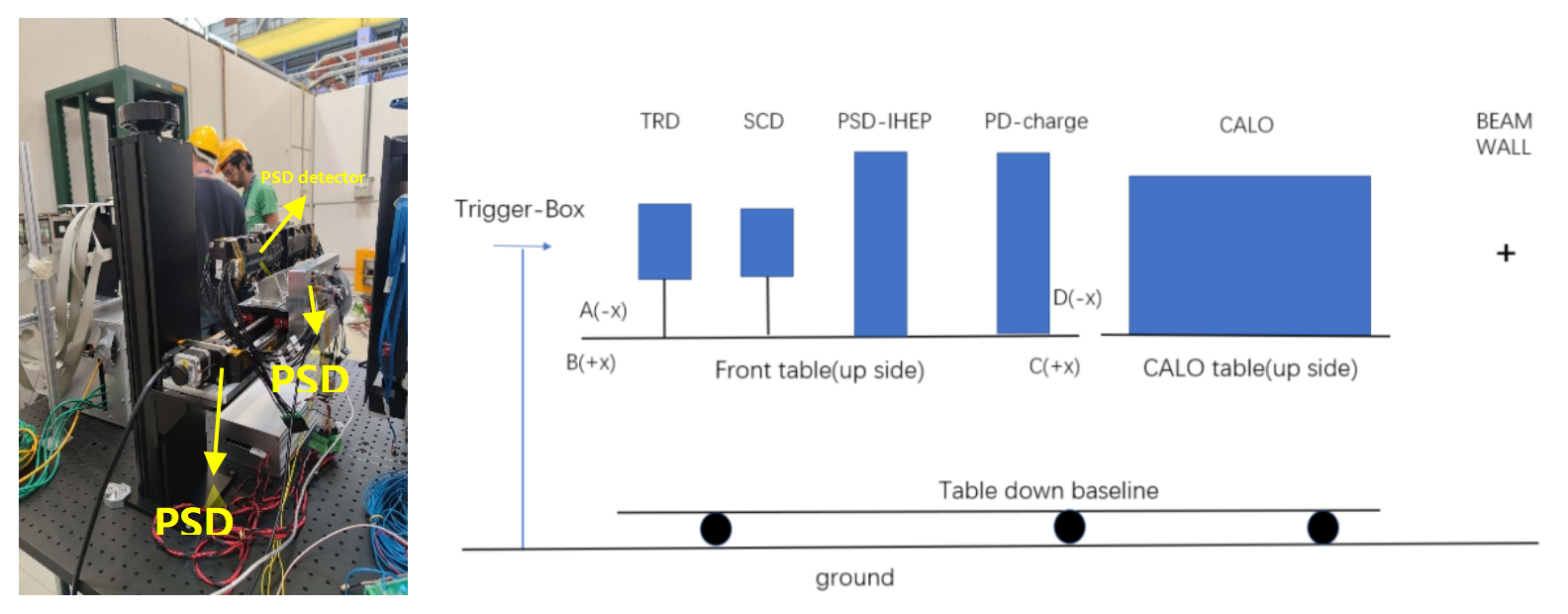}	
	\caption{Layout of the PSD beam test prototype and the SPS-proton beam test prototype} 
	\label{fig3}%
\end{figure}

The PSD prototype, along with other HERD sub-detectors and some auxiliary detectors, participated in the PS, SPS proton, and SPS heavy-ion beam tests. The geometrical layout of each prototype on the SPS proton beamline is shown in Figure \ref{fig3}. The PSD prototype conducted detailed transverse position scanning tests on this beamline. The scanning range was from \SI{-150}{\milli\meter} to \SI{150}{\milli\meter} as shown in Figure \ref{fig2}, with a spacing of \SI{15}{\milli\meter}
 between each position point. During the entire scanning test, the beam particles were mainly electrons (\SI{100}{\giga\electronvolt}, \SI{150}{\giga\electronvolt}, \SI{200}{\giga\electronvolt}, \SI{250}{\giga\electronvolt}), muons (\SI{150}{\giga\electronvolt}, \SI{250}{\giga\electronvolt}), or protons (\SI{300}{\giga\electronvolt}), with relatively large beam spot diameters at some position points. This paper mainly focuses on the research of position reconstruction algorithms using these beam data.

\section{Data preprocessing}
\label{chapter 3}
Given that the test data of the two plastic scintillator units of the PSD prototype are basically consistent and the small-sized SiPMs cannot effectively detect MIPs signals, this paper only uses the data from the 6 large-sized SiPM channels of the first detector, namely the three channels on the inner side of the left and right ends, as shown in Figure \ref{fig2}. The position information of each beam event is obtained from the trajectory detector SCD-it (Italian SCD detector) that participated in the beam test simultaneously. By combining with SCD-it, the hit position on the PSD of the events triggered jointly with PSD can be inferred from its reconstructed trajectory information and geometrical layout, and this position is considered as the "true" position of the event. Due to the environmental background and the quality of trajectory reconstruction, the raw beam data from PSD needs to undergo some processing and filtering before being used for event position reconstruction. The entire filtering process is mainly divided into the following steps.

Step 1: For the PSD data, it is required that the data from all 6 SiPM channels exceed the threshold to eliminate accidental coincidences caused by noise or environmental background. Then, based on the baseline calibration results of each channel during the test, the ADC value of each channel is subtracted by the corresponding baseline.

\begin{figure}
	\centering 
	\includegraphics[width=0.5\textwidth, angle=0]{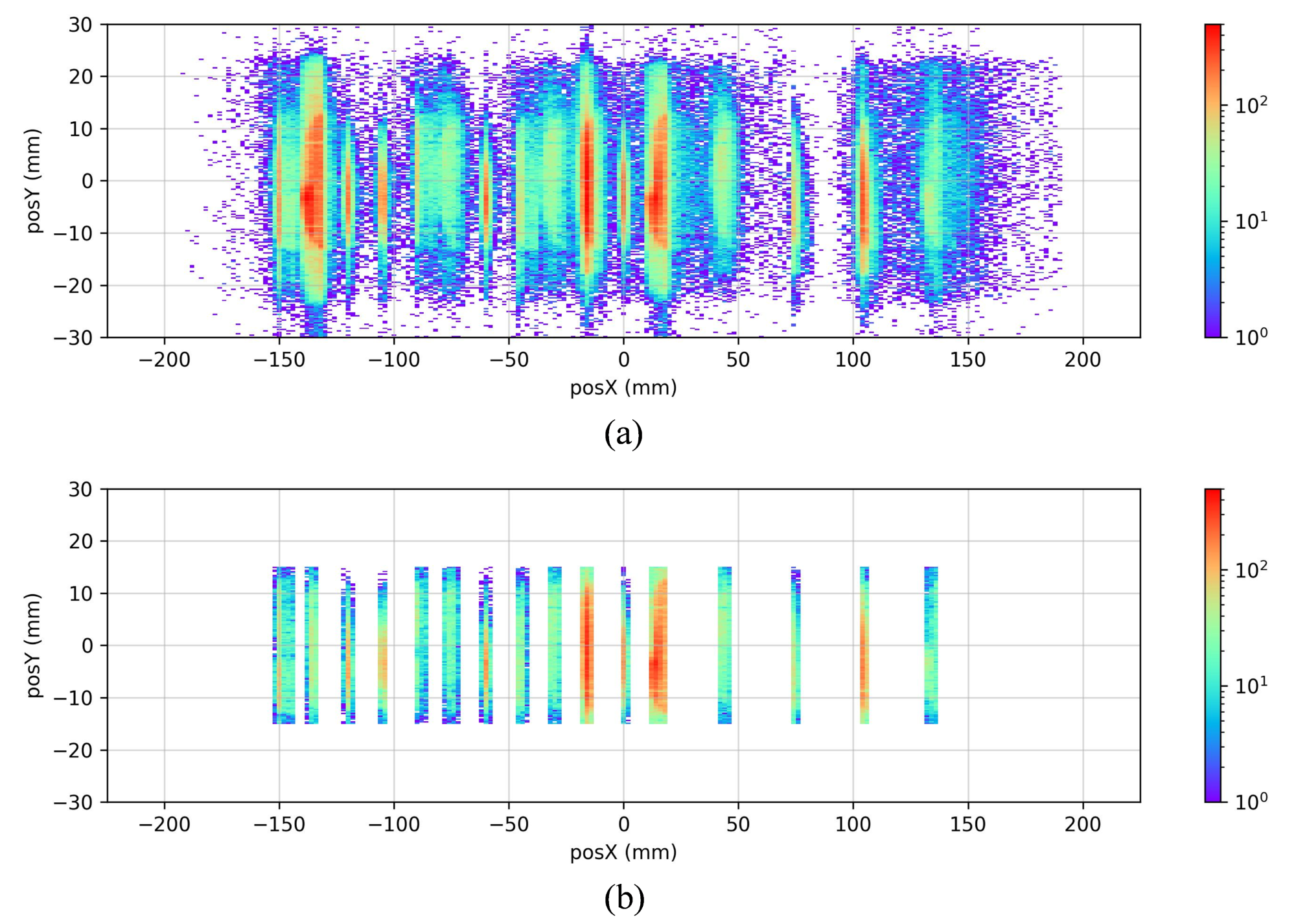}	
	\caption{Distribution of beam events on the PSD platform. (a) Original event distribution, (b) Cleaned event distribution} 
	\label{fig4}%
\end{figure}

Step 2: Based on the trajectory reconstruction parameters of SCD-it, including Chi2 and slope, select events with better trajectory reconstruction quality. Then, filter out events triggered jointly by PSD and SCD-it based on the event trigger number, as shown in Figure \ref{fig4}a.

Step 3: Due to the trapezoidal cross-section of the PSD crystal (with a top width of \SI{30}{\milli\meter} and a bottom width of \SI{50}{\milli\meter}), beam events hitting the edges of the crystal exhibit different fluorescence attenuation effects compared to events hitting the center. Therefore, this paper only uses events from the center of the PSD (i.e., events within the range of $\pm \SI{15}{\milli\meter}$ in the Y direction).

Step 4: Remove beam events that may be caused by multiple particle events or other abnormal situations, as well as events with inaccurate position information calculated by SCD-it. In this paper, some abnormal beam events are excluded by calculating the ADC ratio of the left and right channels on the PSD prototype and the energy spectra of each channel. The ADC ratio at the approximate beam center position shows a Gaussian distribution, and abnormal events beyond the 2 sigma of the Gaussian distribution are removed. For the energy spectra, abnormal events are removed by setting thresholds. In addition, as shown in Figure \ref{fig4}a, the calculated event positions are mostly located near the beam center, but there are a few events that deviate significantly. These abnormal events are considered to have inaccurate position information calculated by SCD-it and need to be removed.

After the screening process described above, a dataset for beam event position reconstruction in this paper was constructed, with a total of approximately 200,000 events, as shown in Figure \ref{fig4}b. The dataset was divided into training set (120,000 events), validation set (40,000 events), and test set (40,000 events) in a ratio of 6:2:2. The training set and validation set were used for training and validating the MLP neural network and the residual KAN neural network models, while the test set was used to evaluate the performance of beam event position reconstruction, including traditional event position reconstruction based on the ratio of the two ends and event position reconstruction based on the residual KAN neural network model.

\section{Beam event position reconstruction based on the ratio of the two ends}
\subsection{Algorithm introduction}

Due to the attenuation effect of fluorescence propagation in the plastic scintillator crystal, when beam particles hit the PSD, the weaker the signal collected by the SiPM, the farther the distance from the SiPM readout position to the hit position of the particle\citep{knoll2010radiation}\citep{2020Fluorescence}. The overall signal strength exhibits an exponential decay trend with distance. For the two symmetric SiPM channels on the PSD, the ADC ratio after subtracting the baseline can be calculated. As the beam position scans from the left SiPM position to the right SiPM position, a monotonically decaying curve of the ADC ratio is formed. By fitting this decay curve with the corresponding formula, a curve relating the ADC ratio to the beam event position information can be obtained. For any beam event, calculating the ADC ratio at its two ends and using this fitted curve, the position of the event on the PSD detector can be determined.

\subsection{Implementation of beam event position reconstruction based on the ratio of the two ends}

In the SPS proton beam test, the PSD has 6 SiPM channels, three on each side. The three channels on the left side are located at \SI{-150}{\milli\meter}, \SI{-120}{\milli\meter}, and \SI{-90}{\milli\meter} of the PSD detector, named Ch1, Ch2, and Ch3, respectively. The three channels on the right side are located at \SI{150}{\milli\meter}, \SI{120}{\milli\meter}, and \SI{90}{\milli\meter}, named Ch11, Ch12, and Ch13, respectively.

\begin{figure*}[!htb]
	\centering 
	\includegraphics[width=0.9\textwidth, angle=0]{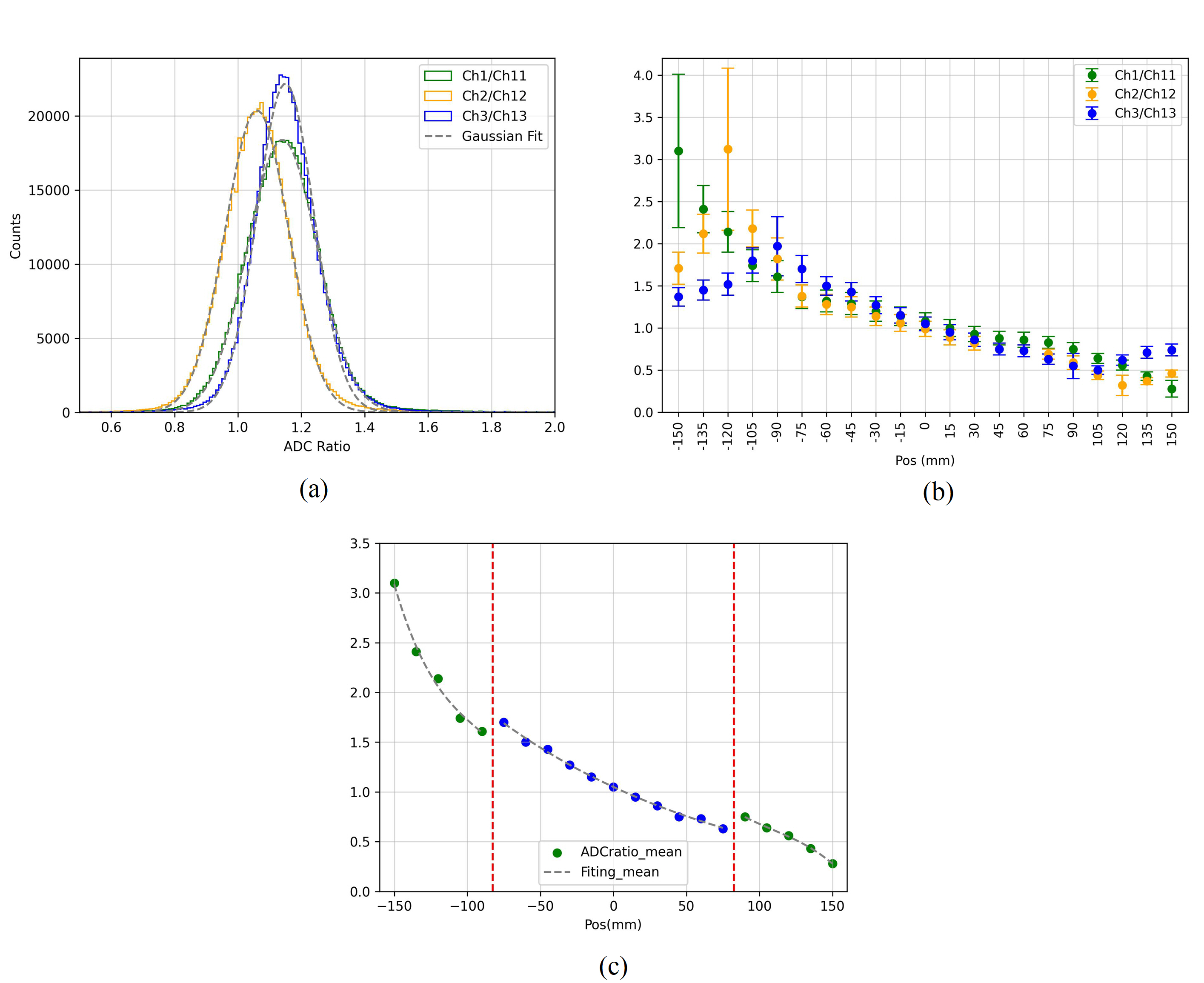}	
	\caption{(a) ADC ratio distribution, (b) Error bars of the ADC ratio at different beam positions, (c) Segmented fitting of the ADC mean} 
	\label{fig5}%
\end{figure*}

First, we calculated the ADC ratios of the three sets of SiPMs at different beam positions (Ch1/Ch11, Ch2/Ch12, and Ch3/Ch13) for the beam events. Figure \ref{fig5}(a) shows the distribution of the ratios when the beam is incident at \SI{-15}{\milli\meter} on the PSD platform. They exhibit Gaussian distributions, and Gaussian fitting was performed to obtain their means and standard deviations. By repeating the same operation for 21 central positions of the beam on the PSD platform, the variation of the ADC ratios with beam position was obtained, as shown in Figure \ref{fig5}(b). When the beam is located in the middle of a set of SiPMs, the ADC ratio of that set shows a decaying trend; when the beam is outside the SiPMs, the trend is opposite. This also indicates that the position reconstruction based on the ADC ratio of the two ends is only applicable between the two ends' SiPMs.

In order to more effectively utilize the decaying trend of the ADC ratios in each segment, and to improve the accuracy and reliability of the position reconstruction, this study adopts a segmented fitting approach for the decay curve of the ADC ratios. In different position segments, we choose the curve sections with higher slopes from Figure \ref{fig5}(b) for fitting, as a higher slope indicates a stronger resolution capability at different positions.

In the range from \SI{-150}{\milli\meter} to \SI{-90}{\milli\meter} and from \SI{90}{\milli\meter} to \SI{150}{\milli\meter}, a third-order polynomial fitting is performed on the decay curve of the ADC ratio of Ch1/Ch11 from Figure \ref{fig5}(b). The fitting formula is as follows:
\begin{equation}
    f(x)=ax^3+bx^2+cx+d
\label{equ1}
\end{equation}
Where  ${a}$,  ${b}$,  ${c}$, and  ${d}$ are constants, and ${a}$ is not equal to 0.

In the range from \SI{-75}{\milli\meter} to \SI{75}{\milli\meter}, an exponential decay fitting is performed on the decay curve of the ADC ratio of Ch3/Ch13 from Figure \ref{fig5}(b). The fitting formula is as follows:
\begin{equation}
    f(x)=ae^{-\frac{x}{t}}+b
\label{equ2}
\end{equation}
Where  ${a}$,  ${t}$, and  ${b}$ are constants, and  ${t}$ is not equal to 0.

\begin{table*}[!ht]
\centering
\begin{tabularx}{\textwidth}{@{\hspace{15pt}}c@{\hspace{60pt}}l@{\hspace{40pt}}l@{\hspace{40pt}}l@{\hspace{40pt}}l}
\hline
Fitting range             & \multicolumn{4}{c}{Fitting parameters}                        \\ \hline
\SI{-150}{\milli\meter} $\sim$ \SI{-90}{\milli\meter} & a=-3.703705e-06 & b=-0.001019 & c=-0.106071 & d=-2.395717 \\
\SI{-75}{\milli\meter} $\sim$ \SI{75}{\milli\meter}   & a=1.15          & t=169.28    & b=-0.10      &             \\
\SI{90}{\milli\meter} $\sim$ \SI{150}{\milli\meter}   & a=-1.234567e-06 & b=0.000403  & c=-0.050151 & d=2.896284  \\ 
\hline
\end{tabularx}
\caption{Fitting parameters for different ranges} 
\label{Table1}
\end{table*}

Figure \ref{fig5}(c) shows the segmented fitting results of the ADC mean values, and Table \ref{Table1} lists the parameters of the segmented fitting formula. For any beam event incident on the PSD platform, the ADC values detected by the 6 channels are used to calculate the corresponding ratios. These ratios are then substituted into the fitting formula mentioned above to determine the specific position of the event on the PSD detector, thus completing the position reconstruction.

\section{Position reconstruction of beam events based on the residual KAN neural network}
\subsection{Overview}

The Multilayer Perceptron (MLP) is a classic neural network model composed of multiple layers of neurons\citep{Taud2018}. Its structure and functionality make it a key component of deep learning, excelling in various tasks such as image classification, text classification, prediction, and regression\citep{2019Deep}. In contrast, the Kolmogorov-Arnold (KAN) network is a promising alternative to MLP\citep{liu2024kankolmogorovarnoldnetworks}. Unlike MLP, which employs fixed activation functions at the nodes ("neurons"), KAN introduces learnable activation functions on the edges ("weights"). The nodes in KAN simply sum the input signals without applying any non-linear transformations. Although this change may seem minor, KAN outperforms MLP in terms of accuracy and interpretability. Residual connections are a widely used technique in deep neural networks\citep{7780459}. In residual connections, the input signal passes through one or more layers of the network and is then added to (or concatenated with) the original input to produce the output. This mechanism enables the network to more easily learn identity mappings, thereby improving gradient flow, reducing the vanishing gradient problem during training, enhancing network performance, and accelerating convergence\citep{2017Image}.

This paper constructs a KAN and MLP neural network model with a residual structure based on the latest KAN network layer to achieve the position reconstruction of beam events on the PSD. The reconstruction results are compared with the traditional algorithms described in Chapter \ref{chapter 3} and the position reconstruction results of the MLP network.

\subsection{Construction and training of the PSD-KAN model}
This paper designs a deep neural network model called PSD-KAN, which contains 8 layers and over 640k parameters, tailored to the characteristics of beam event data. The model takes as input the ADC values of the 6 channels of SIPM on the PSD after subtracting the baseline, and directly outputs the reconstructed position of the beam event on the PSD. Compared to shallow architectures like fully connected networks used in previous studies, the key feature of this model architecture is the introduction of the residual KAN module. The KAN structure replaces fixed activation functions with learnable functions, fundamentally eliminating the dependence on linear weight matrices. By applying the residual KAN module to the position reconstruction of beam events, it can comprehensively extract the characteristics of beam events at different positions on the PSD while effectively preventing issues like gradient vanishing, and enhancing the network's generalization capability.

\begin{figure}
	\centering 
	\includegraphics[width=0.3\textwidth, angle=0]{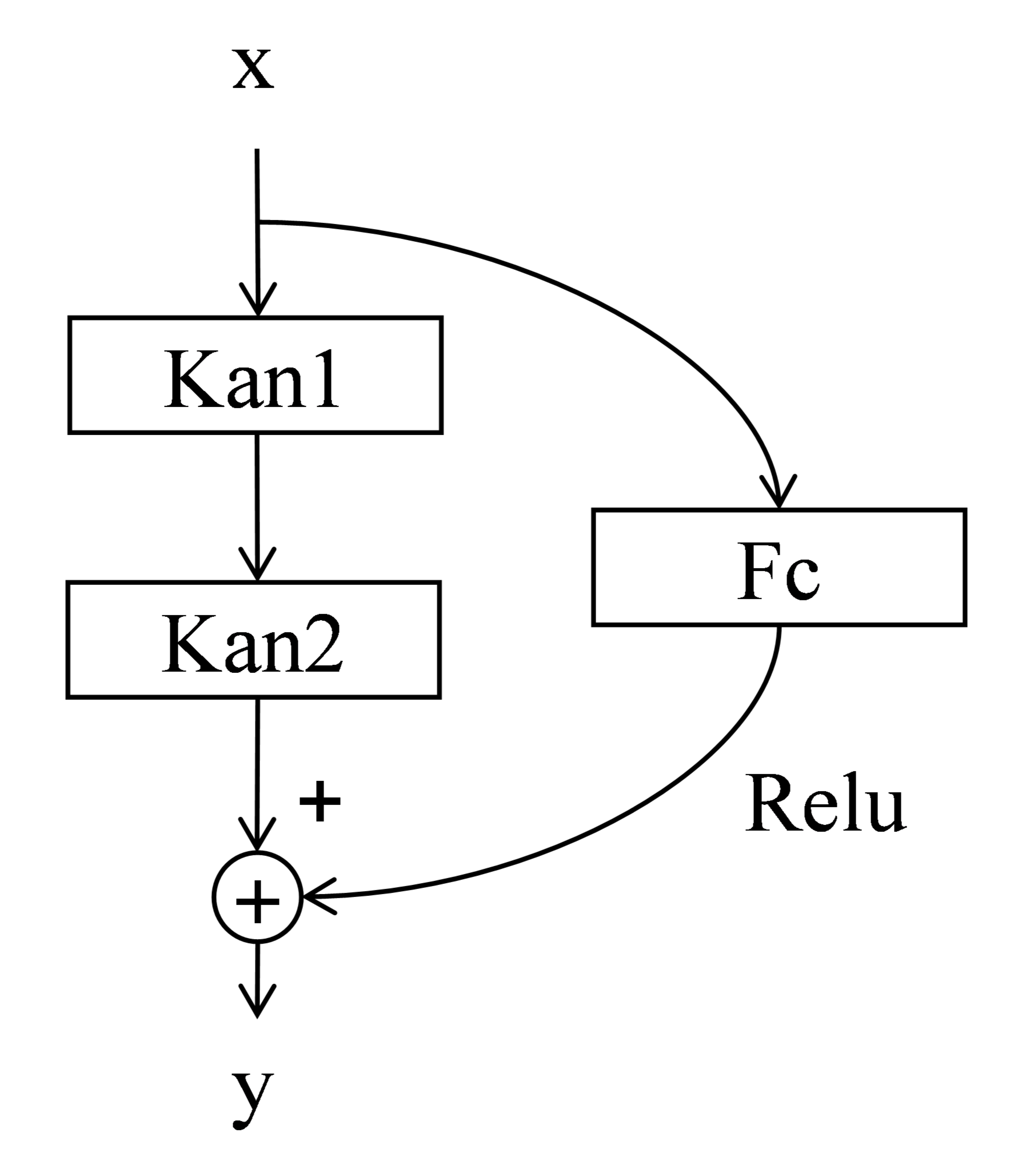}	
	\caption{Residual Network Block R} 
	\label{fig6}%
\end{figure}

\begin{figure*}[!htb]
	\centering 
	\includegraphics[width=0.9\textwidth, angle=0]{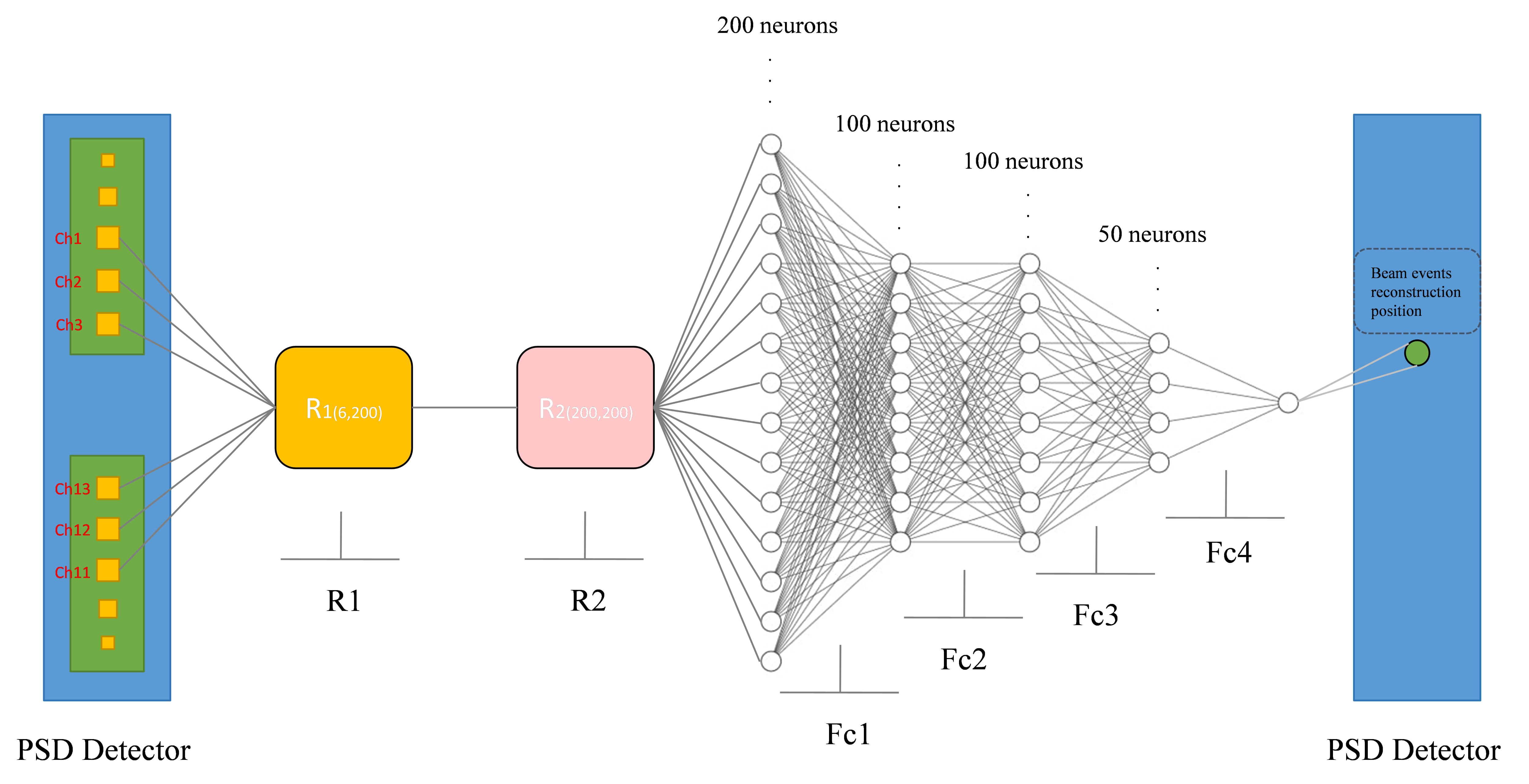}	
	\caption{PSD-KAN model} 
	\label{fig7}%
\end{figure*}

The PSD-KAN neural network model consists of two residual KAN blocks paired with MLP layers. Firstly, the composition of the residual block is shown in Figure \ref{fig6}, denoted by R. It contains two KAN layers, the first KAN layer raises the input dimension to 100 dimensions, and the second KAN layer changes the dimension from 100 dimensions to the output dimension. For the residual connection, the fully connected layer with the Relu activation function is used, the dimension changes from the input dimension to the output dimension, and finally, the residual connection and the KAN layer are dimensionally summed up. The structure of the PSD-KAN model is shown in Figure \ref{fig7}. The input beam event passes through two residual blocks R1 and R2, increasing the dimension from 6 to 200. It then goes through four fully connected layers, each accompanied by a Leaky\_relu activation function, ultimately reducing the dimension to 1, outputting the reconstructed position information of the beam event on the PSD. The PSD-KAN model designed in this paper is the most effective one based on repeated adjustments and experiments. The detailed parameters of each layer in its model architecture are shown in Table \ref{Table2}.

\begin{table}[!ht]
\centering
\setlength{\tabcolsep}{20pt} 
\begin{tabular}{lcc}
\hline
Layer & In\_channels & out\_channels \\  \hline
R1    & 6            & 200           \\
R2    & 200          & 200           \\
Fc1   & 200          & 100           \\
Fc2   & 100          & 100           \\
Fc3   & 100          & 50            \\
Fc4   & 50           & 1             \\
\hline
\end{tabular}
\caption{Parameters of the residual PSD-KAN model layers} 
\label{Table2}
\end{table}

During model training, we use the mean square error (MSE) loss function to evaluate the performance of the model.

\begin{equation}
    MSE = \frac{1}{n} \sum_{i=1}^{n}(y_i - \hat{y_i})^2
\label{equ3}
\end{equation}
Where the variable ${n}$ denotes the sample size, ${y_i}$ is the actual value of the sample, and $\hat{y_i}$ is the predicted value of the model.$(y_i - \hat{y_i})^2$ is defined as the sum of squares of the difference (error) between the true value and the predicted value of each sample.$\sum_{i=1}^{n}$ denotes the summation of the sum of squares of the errors for all samples. In training, we calculate the loss function values of the model on both the training set and the validation set and use the lowest value of the loss function on the validation set as a benchmark to continuously update and retain the parameters of the PSD-KAN model.

\begin{figure}[!ht]
	\centering 
	\includegraphics[width=0.35\textwidth, angle=0]{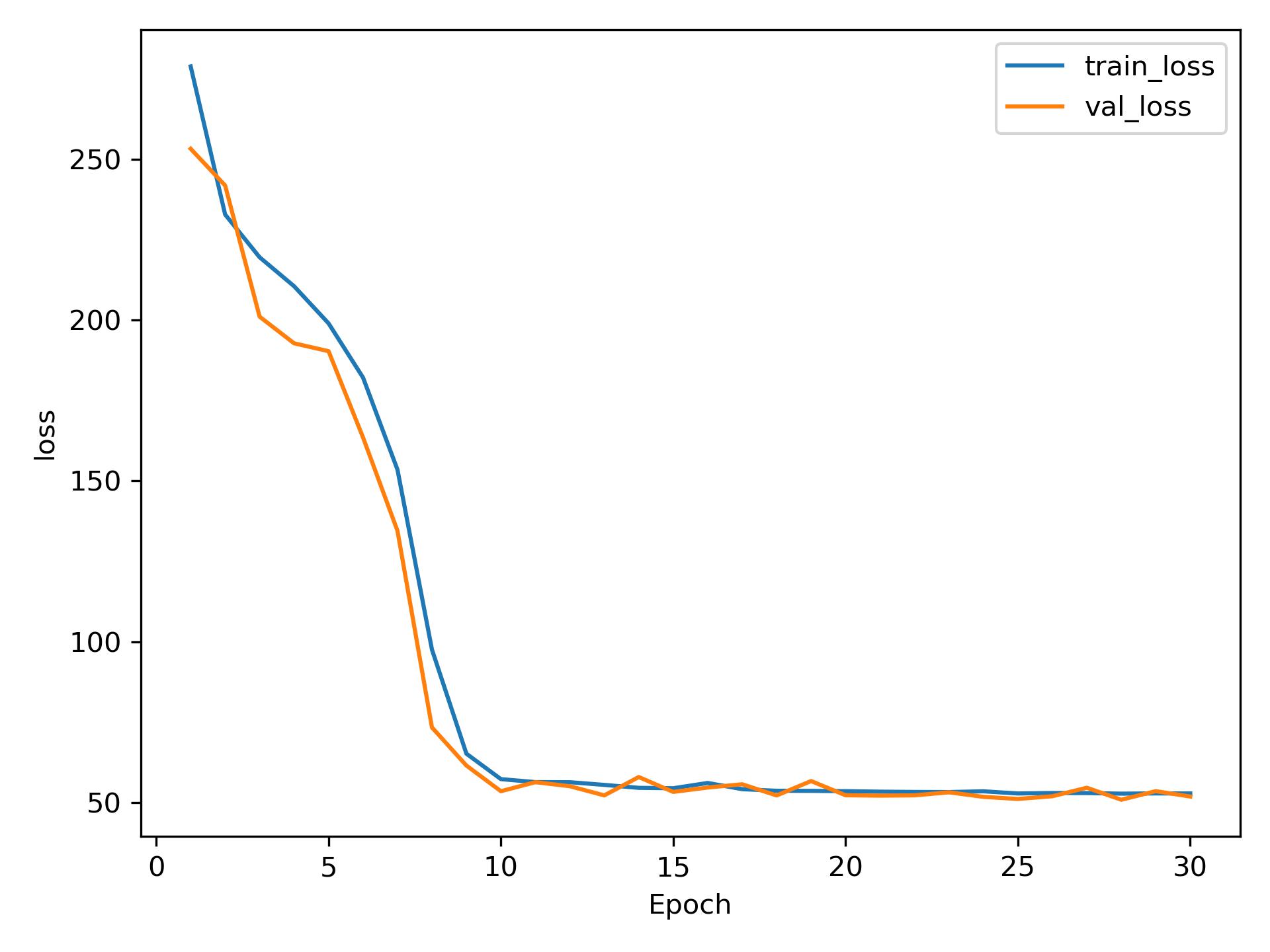}	
	\caption{Loss function during training} 
	\label{fig8}%
\end{figure}

The model training framework is constructed using PyTorch, each batch contains 64 samples, Adam (Adaptive Matrix Estimation) is chosen as the optimization algorithm, and the initial learning rate is set to 0.001. The model training is carried out for a total of 30 cycles to ensure that the model adequately learns the data features and converges to the optimal parameters. In this paper, the preprocessed training set is used to train the PSD-KAN model, and the validation set is used to validate the model in real-time, when the model is trained, the model is finally tested using the test set to determine the model's location reconstruction effect. Figure \ref{fig8} shows the trend of the loss function change on the training set and validation set for the model during the training process, where the lowest value of the loss function is reached on the validation set during the 28th epoch, Therefore, the parameter corresponding to this epoch is taken as the final parameter of the PSD-KAN model.

\subsection{Reconstruction results}
\begin{figure}
	\centering 
	\includegraphics[width=0.44\textwidth, angle=0]{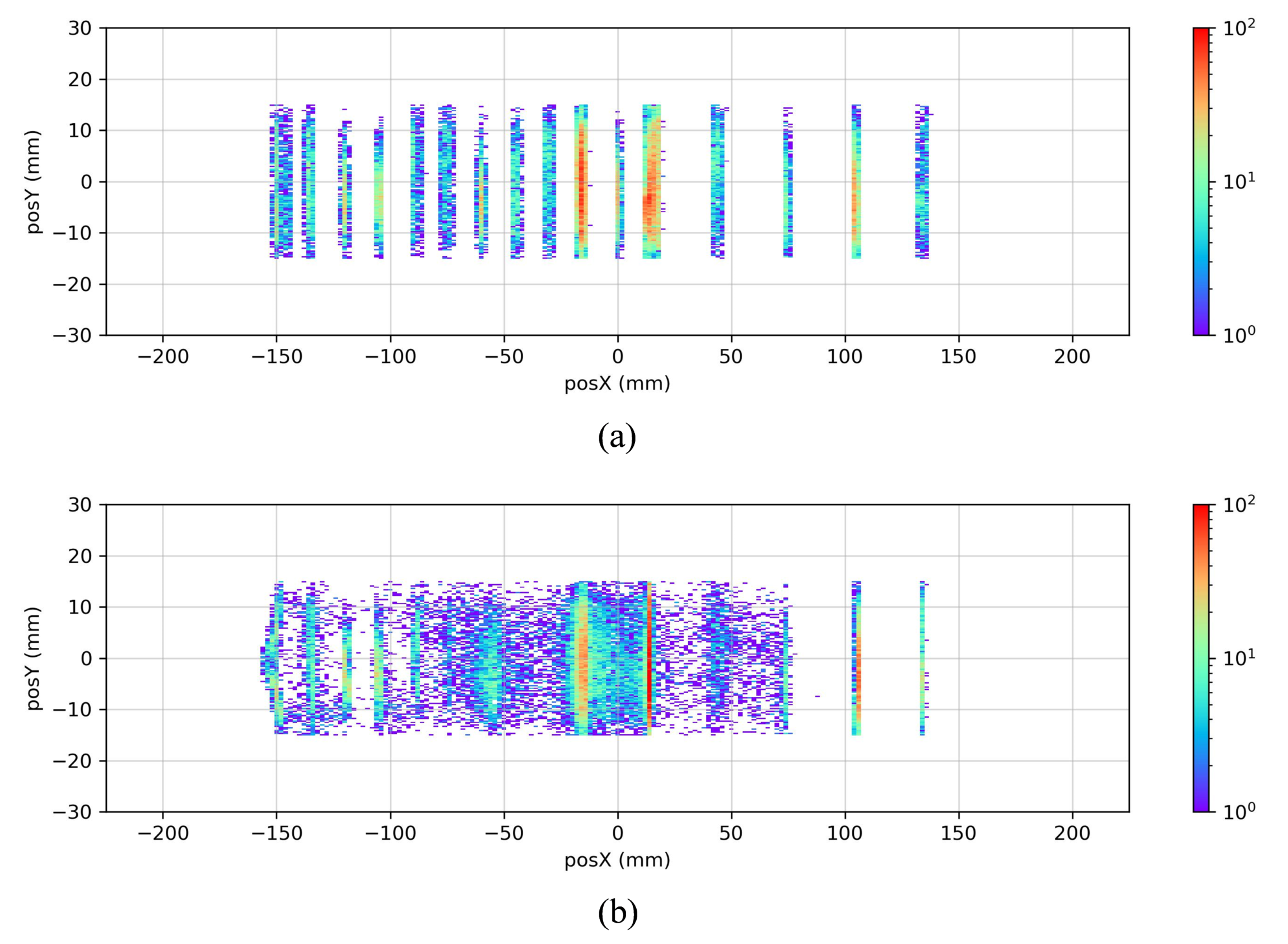}	
	\caption{Positional distribution of beam events on the PSD. (a)the actual position of the event, (b)the reconstructed position of the event} 
	\label{fig9}%
\end{figure}

After completing the training of the PSD-KAN model, the test set of beam instances is input into the trained PSD-KAN model for position reconstruction and is compared with the actual positions of the instances. Figure \ref{fig9}(a) shows the actual position distribution of the beam instances on the PSD detector, and Figure \ref{fig9}(b) shows the reconstructed position distribution of the beam instances.

\subsection{Comparison of reconstruction results}
\begin{figure}
	\centering 
	\includegraphics[width=0.4\textwidth, angle=0]{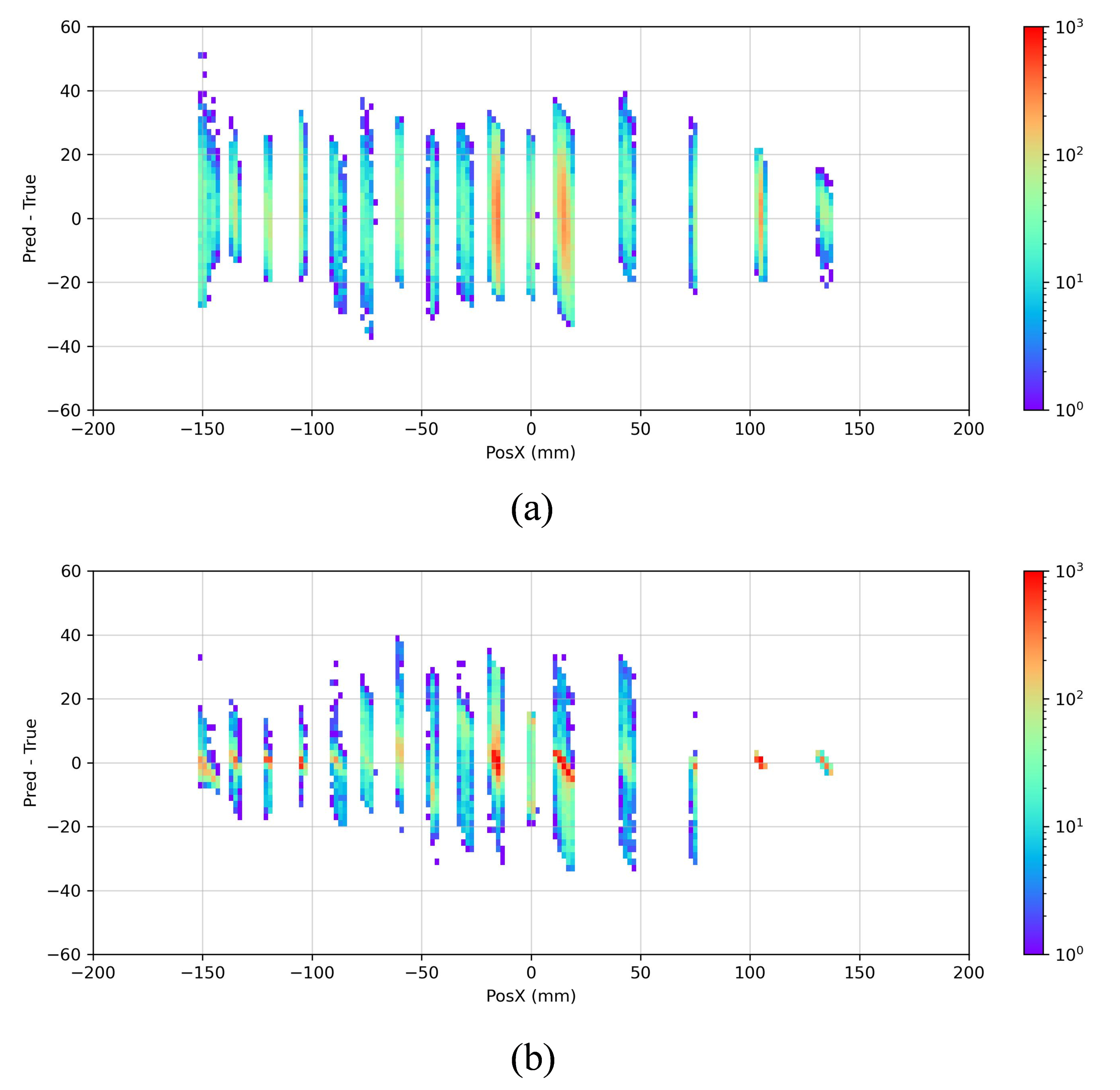}	
	\caption{Reconstruction bias distributions of beam events on PSD. (a) the two-end ratio method, (b) the deep learning method} 
	\label{fig10}%
\end{figure}

The distribution of the deviation of the reconstruction results from the "true" position for the two algorithms is shown in Figure \ref{fig10}. The horizontal axis is the true position in the x-direction of each event, and the vertical axis is the deviation of the reconstructed position from the true position in the x-direction. The distribution of the deviation of the two-end ratio method is shown in Figure \ref{fig10}(a) shows, that the overall distribution range is wide, the reason is that the data points in Figure \ref{fig5} are the average value of ADC ratio at different locations, which cannot represent the situation of all events at that location, which is also a systematic error based on the two-end ratio algorithm to go to do the position reconstruction. Figure \ref{fig10}(b) shows the distribution of deviations based on the PSD-KAN modeling method, the reconstructed position deviation is small, but some events near the middle position of the PSD detector have a large reconstructed position deviation.

\begin{figure}[!ht]
	\centering 
	\includegraphics[width=0.45\textwidth, angle=0]{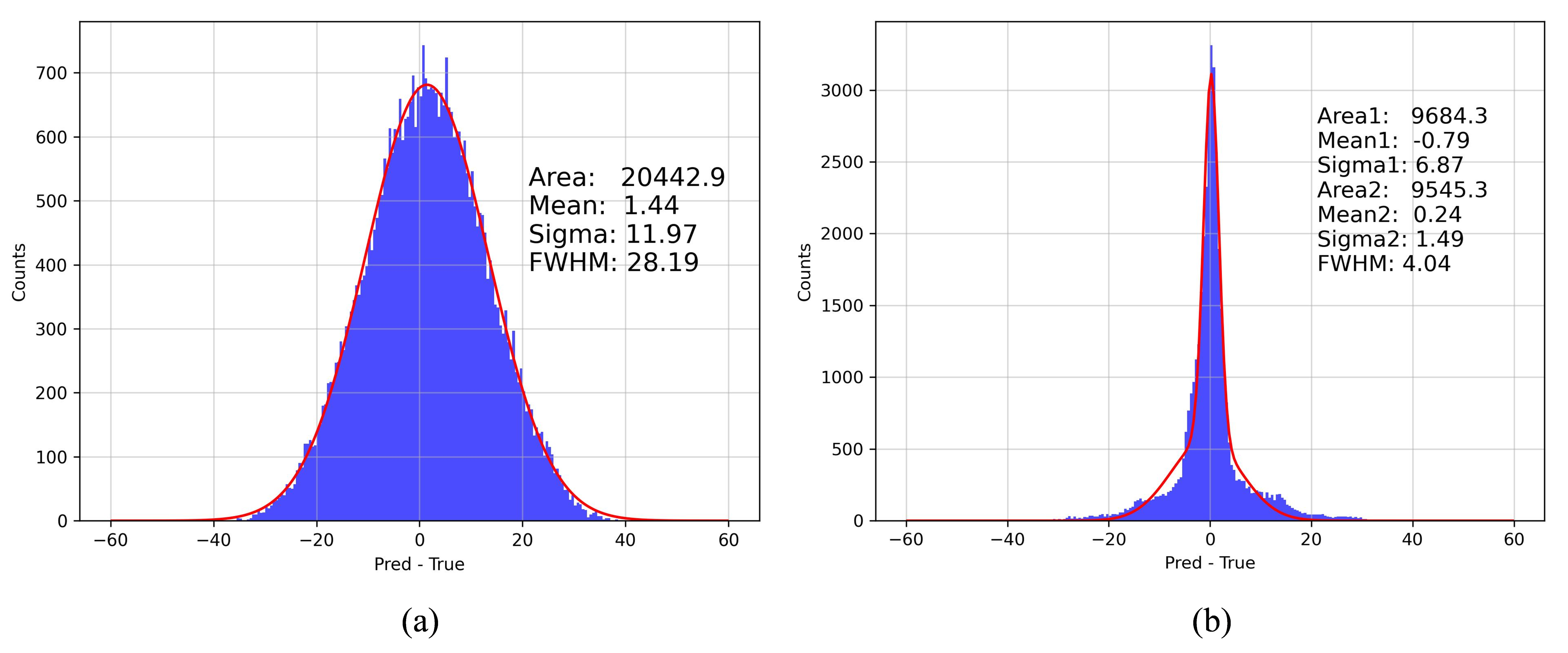}	
	\caption{Comparison of reconstruction bias of event position. (a) the two-end ratio method, (b) the deep learning method} 
	\label{fig11}%
\end{figure}

If the data of all positions are counted, the distributions of the position reconstruction deviations of the two algorithms are shown in Figure \ref{fig11}, and both of them show approximate Gaussian distributions. Among them, Figure \ref{fig11}a shows the result of the position reconstruction deviation of the two-end ratio method, and its FWHM value is 28.19 calculated by combining the Gaussian fitting; Figure \ref{fig11}b shows the result of position reconstruction deviation of PSD-KAN model-based method, and its FWHM value is 4.04 calculated by combining the double Gaussian fitting.

In addition, we have also done beam events position reconstruction using only the MLP network (not mentioned earlier in this paper) whose results are intermediate between the two-end ratio and the PSD-KAN model. Table \ref{Table3} below records the deviation results of these three algorithms for the reconstruction of the positions of beam events, in which the PSD-KAN model designed in this paper is significantly better than the other two methods for the reconstruction of the positions of beam events on the PSD.

\begin{table}[!ht]
\centering
\setlength{\tabcolsep}{9pt} 
\begin{tabular}{lccc}
\hline
deviation & ratio of two ends & MLP     & PSD-KAN \\ \hline
\SI{\pm5}{\milli\meter}      & 32.20\%           & 72.86\% & 73.94\% \\
\SI{\pm7.5}{\milli\meter}    & 47.02\%           & 80.08\% & 80.98\% \\
\SI{\pm10}{\milli\meter}     & 59.59\%           & 85.23\% & 85.85\% \\
\SI{\pm15}{\milli\meter}     & 78.85\%           & 93.33\% & 93.68\% \\
fwhm      & 28.19             & 5.01    & 4.04    \\
\hline
\end{tabular}
\caption{Position reconstruction deviation results } 
\label{Table3}
\end{table}

\begin{figure}[!htb]
	\centering 
	\includegraphics[width=0.45\textwidth, angle=0]{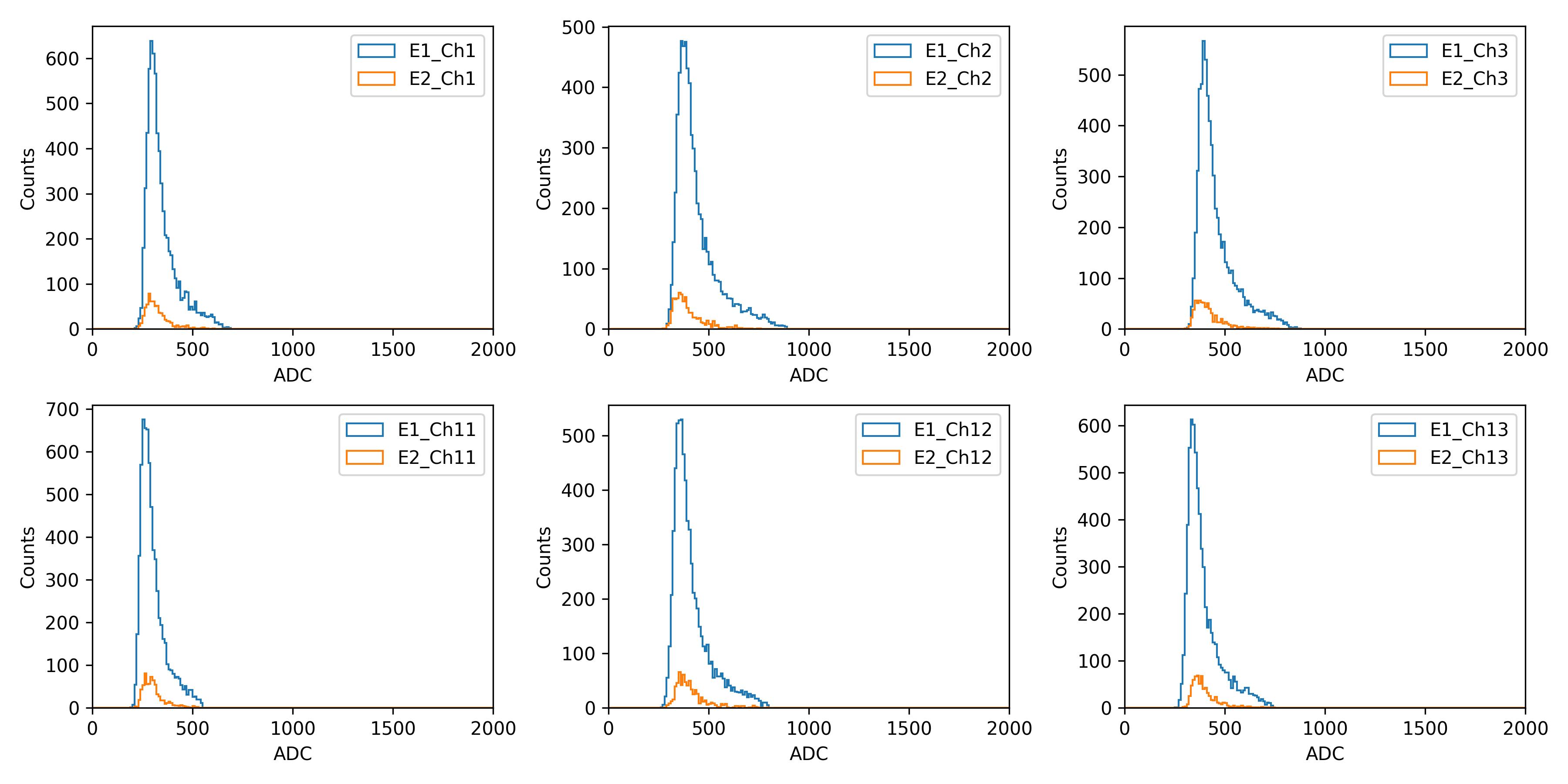}	
	\caption{SIPM 6-channel energy spectra} 
	\label{fig12}%
\end{figure}

\begin{figure}[!htb]
	\centering 
	\includegraphics[width=0.45\textwidth, angle=0]{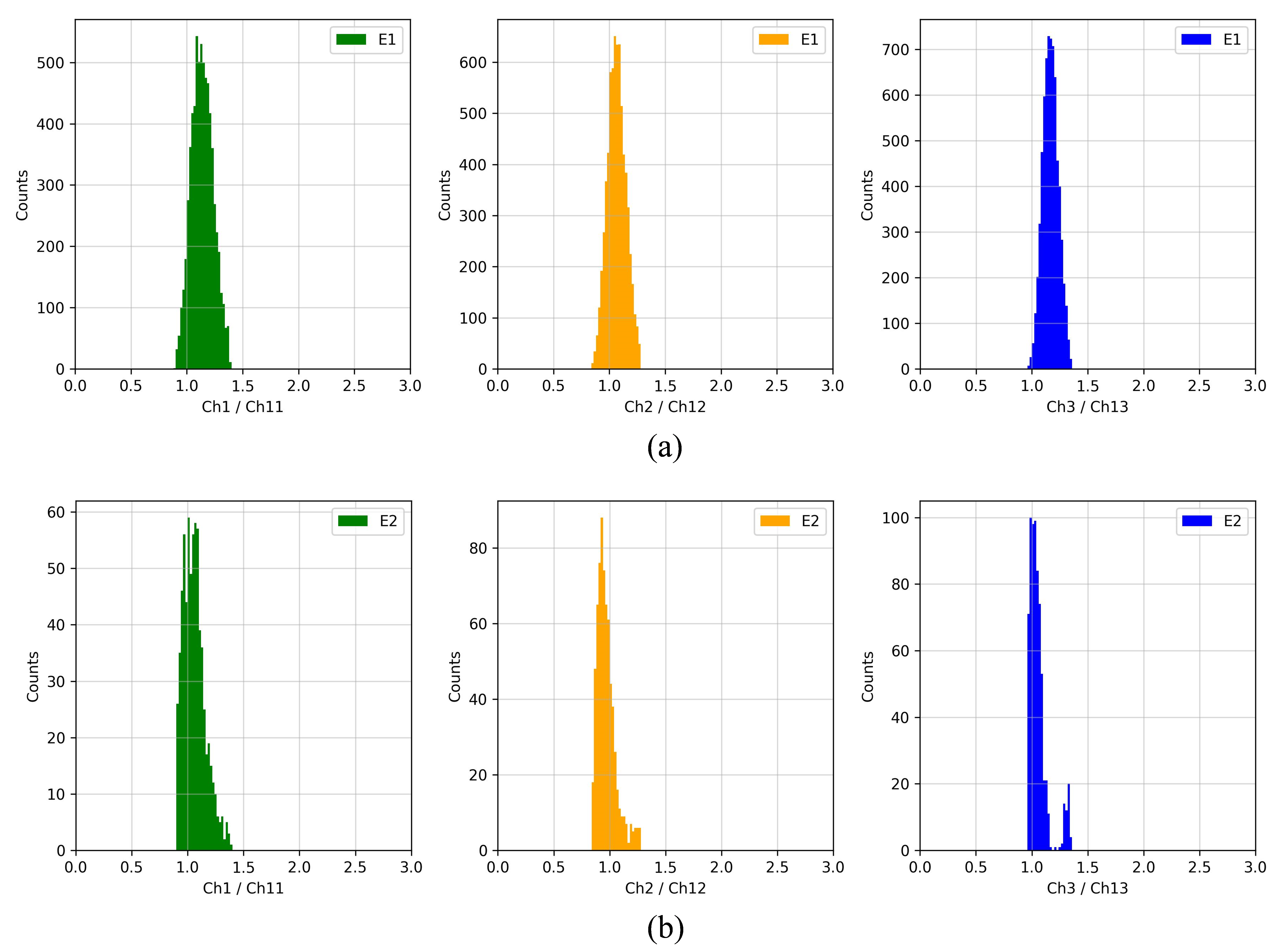}	
	\caption{ADC ratios at both ends. (a) E1 event (position reconstruction deviation within \SI{5}{\milli\meter}), (b) E2 event (position reconstruction deviation outside \SI{15}{\milli\meter})} 
	\label{fig13}%
\end{figure}

To further analyze the reasons for the large reconstruction deviations in some events and the characteristics of these events, the events with reconstruction deviations within $\pm \SI{5}{\milli\meter}$ and outside $\pm \SI{15}{\milli\meter}$, denoted by E1 and E2, respectively, are selected according to the results of Figure \ref{fig10}b. Taking the beam influx position at \SI{-15}{\milli\meter} as an example, Figure \ref{fig12} shows the 6-channel energy spectra of E1 and E2 at this position, from which it can be seen that the energy spectra of the right 3-channel Ch11, Ch12, and Ch13 are basically the same in the distribution of E1 and E2, whereas the energy spectra of the left 3-channel Ch1, Ch2, and Ch3 have a mismatch between the peaks of E1 and E2, and the peaks of E1 at are larger than those of E2. The ADC ratios of the left and right SiPM channels in the case of E1 and E2 are further calculated, as shown in Figure \ref{fig13}. From the figure \ref{fig13}, it is easy to see that the ADC ratio distributions of E1 and E2 show obvious peak position inconsistencies. However, during the model training process, the actual positions given to the two parts of the events of E1 and E2 are close to each other, which leads to a large deviation of the reconstructed position information of the PSD-KAN model for this part of the events of E2. The reason for such events may be that the actual position information calculated from the trajectory detector is incorrect or there is a multi-particle incidence that is not completely removed in the data selection.

\section{Conclusion}
In order to study the HERD PSD position reconstruction algorithm, this paper screens the beam test data of the PSD prototype at CERN and establishes a training set, a validation set, and a test set. Firstly, this paper realizes the position reconstruction based on the traditional algorithm of the two-end ratio. However, the algorithm has certain systematic errors, and the final position reconstruction deviates from the FWHM of about \SI{28}{\milli\meter}. Secondly, this paper builds the PSD-KAN neural network model, and lets the model train on the 6-channel eigenvalues of the beam events and their positional information labels, so as to make the model itself discover the relationship between the inputs and the outputs. The results show that the FWHM of the position reconstruction deviation of the beam events based on the PSD-KAN neural network model is about \SI{4}{\milli\meter} compared to the two-end ratio method, which is significantly better than that of the traditional two-end ratio method. In addition, the SiPM readout position of the HERD PSD is located at the two end surfaces, which are still some distance away from the end surface. The position reconstruction algorithm of the two-end ratio is only applicable to events where the hit position is located between the readout channels. The PSD-KAN neural network model may be more applicable but is limited by the fact that such a location was not scanned for this beam test and could not be verified. At the same time, the deep learning network model is constantly changing, and in the future, there may be better models to apply to event position reconstruction, and we will continue to promote and improve this part of the work.

\bibliographystyle{elsarticle-harv} 
\bibliography{main}


\end{document}